\documentclass[prl,showpacs,amsmath,amssymb,twocolumn, 10pt]{revtex4}

\usepackage{amsthm}
\usepackage{dcolumn}
\usepackage{bm}
\usepackage{graphicx}

\begin{document}

\newtheorem{corollary}{Corollary}
\newtheorem{definition}{Definition}
\newtheorem{example}{Example}
\newtheorem{lemma}{Lemma}
\newtheorem{proposition}{Proposition}
\newtheorem{theorem}{Theorem}
\newtheorem{fact}{Fact}
\newtheorem{property}{Property}
\newcommand{\bra}[1]{\langle #1|}
\newcommand{\ket}[1]{|#1\rangle}
\newcommand{\braket}[3]{\langle #1|#2|#3\rangle}
\newcommand{\ip}[2]{\langle #1|#2\rangle}
\newcommand{\op}[2]{|#1\rangle \langle #2|}

\newcommand{\tr}{{\rm tr}}
\newcommand {\E } {{\mathcal{E}}}
\newcommand {\F } {{\mathcal{F}}}
\newcommand {\diag } {{\rm diag}}

\title{Any $2 \otimes n$ subspace is locally distinguishable}
\author{Nengkun Yu}
\email{nengkunyu@gmail.com}
\author{Runyao Duan}
\email{runyao.duan@uts.edu.au}
\author{Mingsheng Ying}
\email{mying@it.uts.edu.au}

\affiliation{State Key Laboratory of Intelligent Technology and Systems, Tsinghua National Laboratory\protect\\ for Information Science and Technology, Department of Computer Science and Technology,\protect\\ Tsinghua University, Beijing 100084, China}
\affiliation{Centre for Quantum Computation and Intelligent Systems (QCIS), Faculty of
Engineering and Information Technology, University of Technology,
Sydney, NSW 2007, Australia}

\date{\today}

\begin{abstract}
 A subspace of a multipartite Hilbert space is called \textit{locally indistinguishable} if any orthogonal basis of this subspace cannot be perfectly distinguished by local operations and classical communication. Previously it was shown that any $m\otimes n$ bipartite system such that $m>2$ and $n>2$ has a locally indistinguishable subspace. However, it has been an open problem since 2005 whether there is a locally indistinguishable bipartite subspace with a qubit subsystem. We settle this problem by showing that any $2\otimes n$ bipartite subspace is locally distinguishable in the sense it contains a basis perfectly distinguishable by LOCC. As an interesting application, we show that any quantum channel with two Kraus operations has optimal environment-assisted classical capacity.
\end{abstract}

\pacs{03.67.-a, 3.65.Ud}

\maketitle
\textit{1. Introduction:}--LOCC distinguishability of finite set of orthogonal multipartite states is a fundamental task in quantum information theory, it has
attracted much attention and has been extensively studied as it has important implications in  classical data hiding \cite{TDL01} and channel capacity \cite{GW04, HK04,WAT05}.

Orthogonal states can always be exactly distinguished if there are no restrictions on the measurements one can perform. However, the discrimination of multipartite states is difficult when only local operations and classical communication (LOCC) is allowed. Many results on LOCC discrimination seem rather counterintuitive. For instance, Bennett et al discovered that there exist $3\otimes 3$ orthonormal pure product bases that are indistinguishable by LOCC \cite{BDFM+99}. Furthermore, it was shown that the members of an unextendible product basis(UPB) are not perfectly distinguishable by LOCC \cite{BDMS+99}. On the other hand, it has been proven that any two orthogonal multipartite quantum states, no matter entangled or not, can be perfectly distinguished by LOCC \cite{WSHV00}. Some powerful methods for checking distinguishability were introduced in \cite{GKRS+01,HSSH03}.

In 2005 Watrous demonstrated that there
exist a class of $m\otimes m$ subspaces having
no orthonormal bases  locally distinguishable if $m>2$ \cite{WAT05}. Such subspaces are named locally indistinguishable subspaces; otherwise, they are said to be locally distinguishable. Watrous also proved that there is no $2\otimes 2$ locally indistinguishable subspace, by directly employing the results from \cite{WH02}. Winter's result \cite{WIN05} implies that the existence of bipartite subspace $\mathcal{Q}$ such that $\mathcal{Q}^{\otimes k}$ is locally indistinguishable for any $k$. Duan et al generalized Watrous's result to the most general $m\otimes n$ systems for $m\neq n$ and the multipartite setting, and found locally indistinguishable subspaces with smaller dimensions \cite{DFXY09,DXY10}. Most notably, it was shown that any subspace spanned by three-qubit UPB is locally indistinguishable, and there exists a $3$-dimensional three-qubit locally distinguishable subspace. An interesting question remains to be answered is whether there is any $2\otimes n$ locally indistinguishable subspace.

The main contribution of this Letter is to answer the above question negatively. We show that that any  $2\otimes n$ subspace is locally distinguishable. Combining with the previous results \cite{WAT05,DFXY09,DXY10}, we conclude that there is no locally indistinguishable $m\otimes n$ subspace if and only if one of $m$ or $n$ should be $2$. Our key techniques can be used to study the distinguishability of three-dimensional bipartite subspace which contains a product state. We show that any such subspace has a basis that can be distinguished under local projective measurements and one-way classical communication (LPCC).

Since the connection between distinguishability of subspace and capacity of channel, classical corrected capacity of a quantum channel is considered. The classical corrected capacity of channel introduced by Hayden and King is defined as the best classical capacity can be obtained when the receiver of the channel is capable to select an optimal measurement on the channel's output by using classical information obtained from a measurement on the environment \cite{HK04}. This environment-assisted model was first introduced by Gregoratti and Werner in \cite{GW04}, where they are interested in correcting the errors incurred from sending quantum information. According to the well known result by Walgate et al \cite{WSHV00}, Hayden and King were able to show that the classical corrected capacity of any quantum channel is at least one bit \cite{HK04}. In particular, the existence of locally indistinguishable subspaces implies the existence of quantum channel with suboptimal classical corrected capacity, that is, the corrected capacity is less than $\log_2 d$ with $d$ the dimension of the input state space. In contrast, our result signifies  that the classical corrected capacity of any quantum channel with only two Kraus operators is  always optimal.

\textit{2. Main result:}--Suppose there is a $2\otimes n$ quantum system shared by Alice and Bob such that Alice is the owner of the qubit. We will show that any $2\otimes n$ subspace has an orthogonal basis which can be perfectly distinguished by some protocol where Alice goes first \cite{WH02}.

We will make use of the following Lemma from \cite{WH02}, which gives a complete characterization of the local discrimination of orthogonal $2\otimes n$ states when Alice goes first.

\begin{lemma}\label{lemma1}\upshape
If Alice goes first, a set of $k$ $2\otimes n$ orthogonal states $\{\ket{\psi_i}:1\leq i\leq k\}$ is
locally distinguishable if and only if there is an orthonormal basis
$\{\ket{0}, \ket{1}\}_A$ such that:
\begin{equation}\label{restriction}
\ket{\psi_i}=\ket{0}\ket{\eta_0^i}+\ket{1}\ket{\eta_1^i}
\end{equation}
where $\ip{\eta_0^i}{\eta_0^j}=\ip{\eta_1^i}{\eta_1^j}=0$ for all $i\neq j$.
\end{lemma}

Now we are ready to present our main result as follows.
\begin{theorem}\label{theorem1}\upshape
For any $2\otimes n$ subspace $\mathcal{Q}$, there exists orthogonal basis $\{\ket{\phi_i}|1\leq i\leq d\}$ which can be perfectly distinguished by some Alice goes first LOCC protocol, where $d$ is the dimension of $\mathcal{Q}$.
\end{theorem}

\textbf{Proof:}---We only need to show that $\mathcal{Q}$ has orthogonal basis  $\{\ket{\psi_i}|1\leq i\leq d\}$ with the form of Eq.(\ref{restriction}).

Choose arbitrary basis of the qubit's system, called $\{\ket{0}, \ket{1}\}$.

Let $\{\ket{\varphi_i}|1\leq i\leq d\}$ an orthonormal basis of $\mathcal{Q}$ with $\ket{\varphi_i}=\ket{0}\ket{\zeta_0^i}+\ket{1}\ket{\zeta_1^i}$, then an $n$-by-$d$ matrix $A$ can be defined as $A=(\ket{\zeta_0^1},\ket{\zeta_0^2}\cdot\cdot\cdot,\ket{\zeta_0^d})$. From singular value decomposition, there exists a factorization of the form $A=W\Lambda V$, where $W\in U(n),V\in U(d)$ and the matrix $\Lambda$ is $n$-by-$d$ diagonal matrix with nonnegative real numbers on the diagonal, it's worth to notice that $AV^{\dag}$ is a $n$-by-$d$ matrix with orthogonal columns.

For any unitary $U=(u_{ij})_{d\times d}$, we can get another orthonormal basis of $\{\ket{\phi_i}|1\leq i\leq d\}$ such that $$\ket{\phi_i}=\sum_{j=1}^d u_{ji}\ket{\varphi_j}= \ket{0}\ket{\eta_0^i}+\ket{1}\ket{\eta_1^i}.$$
where  $\ket{\eta_0^i}=\sum_{j=1}^d u_{ji}\ket{\zeta_0^j}$ and $\ket{\eta_1^i}=\sum_{j=1}^d u_{ji}\ket{\zeta_1^j}$.The essential piece is that the matrix $A$ becomes $AU=(\ket{\eta_0^i},\ket{\eta_0^i}\cdot\cdot\cdot,\ket{\eta_0^d})$ under this transformation $U$.

Now let $U=V^{\dag}$ be the unitary transformation, then $AU=AV^{\dag}=W\Lambda$ is a matrix with orthogonal columns, which means that after the transformation $U$, the basis $\ket{\phi_i}=\ket{0}\ket{\eta_0^i}+\ket{1}\ket{\eta_1^i}$  satisfy that $\ip{\eta_0^i}{\eta_0^j}=0$ for all $i\neq j$, note that the Orthogonality to $\ket{\phi_i}$ implies $\ip{\eta_0^i}{\eta_0^j}+\ip{\eta_1^i}{\eta_1^j}=0$, so $\ip{\eta_1^i}{\eta_1^j}=0$, thus Eq.(\ref{restriction}) is satisfied, which completes the proof of our theorem. \hfill
$\blacksquare$

According to the proof, one can find that Alice even has the freedom to preselect arbitrary orthonormal basis to be her projective measurement basis, which is to say that for any $\{\ket{0}, \ket{1}\}$, there exists a basis of $\mathcal{Q}$ satisfying Eq.(\ref{restriction}).

Combining our result with the
existence of locally(separability) indistinguishable subspace for $m\otimes n$ when $m,n>2$ \cite{WAT05, DFXY09,DXY10}, we have the following corollary:
\begin{corollary}\label{collary1}\upshape
There exists $m\otimes n$ subspace indistinguishable by LOCC (or LPCC, Separable operations) if and only if $m,n>2$.
\end{corollary}

Also, we can employ the techniques deriving the above theorem  to show that
\begin{corollary}\label{corollary2}\upshape
Any three-dimensional bipartite subspace $\mathcal{Q}$ that contains a product state is one-way LPCC distinguishable.
\end{corollary}

\textbf{Proof:}---Without loss of generality, assume that $\ket{0}\ket{0}\in \mathcal{Q}$, let $\{\ket{\varphi_i}|1\leq i\leq 3\}$ be an orthonormal basis of $\mathcal{Q}$ with $\ket{\varphi_1}=\ket{0}\ket{0}$. Let $\mathcal{P}=\mathrm{span}\{\ket{\varphi_2},\ket{\varphi_3}\}$, similar as the proof of Theorem \ref{theorem1}, one can find orthonormal basis $\ket{\phi_2},\ket{\phi_3}$ of $\mathcal{P}$, such that
\begin{eqnarray*}
\ket{\phi_2}&=&\ket{0}\ket{\eta_0^2}+\sum_{i\neq 0}\ket{i}\ket{\alpha_i},\\
\ket{\phi_3}&=&\ket{0}\ket{\eta_0^3}+\sum_{i\neq 0}\ket{i}\ket{\beta_i},
\end{eqnarray*}
with $\ip{\eta_0^2}{\eta_0^3}=0$. Then  $\{\ket{\phi_i}|1\leq i\leq 3\}$ are the basis of $\mathcal{Q}$.

According to \cite{WSHV00}, one can
always find a orthogonal basis $\{\ket{0}',\ket{1}',\cdot\cdot\cdot,\ket{m}'\}$ of $\mathcal{C}^{m}$ with $\ket{0}=\ket{0}'$, in which the two orthogonal states can be represented
\begin{eqnarray*}
\ket{\phi_2}&=&\ket{0}'\ket{\eta_0^2}+\sum_{i\neq 0}\ket{i}'\ket{\alpha_i}',\\
\ket{\phi_3}&=&\ket{0}'\ket{\eta_0^3}+\sum_{i\neq 0}\ket{i}'\ket{\alpha_i^{\bot}}',
\end{eqnarray*}
where $\ket{\alpha_i'}$ are not normalized, and $\ket{\alpha_i'^{\bot}}$ is orthogonal to $\ket{\alpha_i'}$.

In order to distinguish them, one can perform the projective measurement $\{P_0,P_1,\cdot\cdot\cdot,P_m\}$ upon the first part where $P_i=\op{i'}{i'}$. If the outcome is $P_0$, then the left three mutual orthogonal states are on one part, this leads to perfect discrimination. If the our come is $P_i$ with $i>0$, the left two states are pure orthogonal, they are LPCC distinguishable. $\mathcal{Q}$ is one-way LPCC distinguishable. \hfill
$\blacksquare$

This protocol also works when the other part goes first.

\textit{3. Classical corrected capacity of rank two channel:}--
Any quantum channel $\Phi$ can be regarded as arising from a unitary interaction $U$ of the principle system $\mathcal{H}$ and an environment system $\mathcal{E}$. Without loss of generality, we can let $$\Phi(\rho)=tr_{env}[U(\rho\otimes\op{\varepsilon}{\varepsilon})U^{\dag}].$$
Because $U$ is unitary, it will map the orthogonal input state to orthogonal state in $\mathcal{H}\otimes \mathcal{E}$. However after the trace over of the environment system, the output of the system may be no longer orthogonal. Thus they cannot be distinguished perfectly, which decreases the classical capacity of the channel.

It is possible to enhance channel capacity using measurements
on the environment in addition to measurements on the principal system \cite{HK04,GW04}.

For any rank two channel, the dimension of the environment can be assumed as 2, thus the whole space $\mathcal{H}\otimes \mathcal{E}$ is a $n\otimes 2$ space, before tracing over the the environment, the output space $\mathcal{Q}=U(\mathcal{H}\otimes \ket{\varepsilon})$ is a $d$-dimensional subspace of $\mathcal{H}\otimes \mathcal{E}$, where $d$ is the dimension of $\mathcal{H}$. According to Theorem \ref{theorem1}, it is distinguishable by some environment goes first LPCC protocol, which means that for any orthonormal basis $\{\ket{0},\ket{1}\}$, there is an orthonormal basis $\{\ket{\phi_i}|1\leq i\leq d\}$ of $\mathcal{Q}$ can be represented as Eq.(\ref{restriction}), and thus can be distinguished by some environment goes first LPCC protocol. One can easily verify that the basis $\{\ket{\phi_i}|1\leq i\leq d\}$ corresponds to an input basis $\{\ket{\psi_i}|1\leq i\leq d\}$ of $\mathcal{H}$ with $\ket{\phi_i}=U(\ket{\psi_i}\otimes \ket{\varepsilon})$.
We have therefore proved the following corollary.
\begin{corollary}\label{corollary3}
Any quantum channel with two Kraus operators has optimal environment-assisted classical capacity.
\end{corollary}
\textit{4. Conclusion:}--We have proven that there is no $2\otimes n$ locally indistinguishable subspace. The local distinguishability of such subspace implies that all rank two channel's environment-assisted classical capacity is optimal.

There are several interesting, unanswered questions relating
to the distinguishability of subspaces having. For instance, if there exists three-dimensional indistinguishable multipartite subspace? The tripartite qubit indistinguishable example has been given in \cite{DXY10}, what left is the bipartite case. We have shown that for all three-dimensional subspaces with a product state, the answer is negative \ref{corollary2}. Numerical evidence was presented to show that any three-dimensional subspace of $C^3\otimes C^n$ has an orthonormal basis which can be reliably distinguished using one-way LOCC in \cite{KM05}.
Recall our proof of Theorem \ref{theorem1}, Alice can perform arbitrary
projective measurement, which indicated that the freedom to preselect an orthonormal basis of one part is not used for this case. Is this freedom helpful for subspace discrimination, particularly, is any three-dimensional subspace of $\mathcal{C}^3\otimes\mathcal{C}^3$ LPCC distinguishable?

This work was partly supported by the Natural Science Foundation of
China (Grant Nos.60736011 and 60702080), and QCIS, University of Technology,
Sydney.


\begin{thebibliography}{99}

\bibitem{TDL01} B. M. Terhal, D. P. DiVincenzo, and D. W. Leung, Phys. Rev.
Lett. 86, 5807 (2001).

\bibitem{GW04} M. Gregoratti and R. F. Werner, J. Math. Phys. 45, 2600 (2004).

\bibitem{HK04} P. Hayden and C. King, Quantum Information and Computation \textbf{5}, 156 (2005).

\bibitem{WAT05} J. Watrous, Phys. Rev. Lett. \textbf{95}, 080505 (2005).

\bibitem{WIN05} A. Winter, arxiv: quant-ph/0507045.

\bibitem{BDFM+99} C. H. Bennett, D. P. DiVincenzo, C. A. Fuchs, T. Mor,
E. Rains, P.W. Shor, J.A. Smolin, and W. K. Wootters,
Phys. Rev. A 59, 1070 (1999).

\bibitem{BDMS+99} C. H. Bennett, D. P. DiVincenzo, T. Mor, P.W. Shor, J.A.
Smolin, and B. M. Terhal, Phys. Rev. Lett. 82, 5385 (1999).

\bibitem{WSHV00} J. Walgate, A.J. Short, L. Hardy, and V. Vedral, Phys.
Rev. Lett. 85, 4972 (2000).

\bibitem{GKRS+01} S. Ghosh, G. Kar, A. Roy, A. Sen(De), and U. Sen, Phys.
Rev. Lett. 87, 277902 (2001).

\bibitem{HSSH03} M. Horodecki, A. Sen(De), U. Sen, and K. Horodecki,
Phys. Rev. Lett. 90, 047902 (2003).

\bibitem{WH02} J. Walgate and L. Hardy, Phys. Rev. Lett. 89, 147901
(2002).

\bibitem{DFXY09} R. Duan, Y. Feng, Y. Xin, M. Ying, IEEE Trans. Inform. Theory \textbf{55}, 1320 (2009).

\bibitem{DXY10} R. Duan, Y. Xin, and M. Ying, Phys. Rev. A \textbf{81}, 032329 (2010). Arxiv: quant-ph/0708.3559.

\bibitem{KM05} C. King and D. Matysiak, quant-ph/0510004.
\end{thebibliography}
\end{document}